\documentclass{PoS}
\usepackage{citesort}
\newcommand{\eq}{\begin{eqnarray}}
\newcommand{\en}{\end{eqnarray}}

\title{Three particles on the lattice}

\ShortTitle{Three particles on the lattice}

\author{\speaker{Akaki Rusetsky}\\
HISKP and BCTP, Rheinische Friedrich-Wilhelms Universit\"at Bonn, 53115 Bonn, Germany\\
        E-mail: \email{rusetsky@hiskp.uni-bonn.de}}

      \abstract{I give a review of existing frameworks, which are designed to analyse
        lattice data in the three-particle sector. A particular emphasis is laid on the
        foundations of the theory, where the separation of the short- and long-range effects
        plays a central role. It is shown that the use of the effective field theory approach
        enables one to carry out this separation in a natural and efficient way. In conclusion,
        a few examples of the analysis of lattice data are considered for illustration.
      }

\FullConference{37th International Symposium on Lattice Field Theory - Lattice2019\\
		16-22 June 2019\\
		Wuhan, China}

\begin{document}

\section{Introduction}

The recent interest to the study of three-particle dynamics on the lattice is related
to several important physical problems. First of all, one would like to investigate the properties of the
resonances, which decay exclusively into three-body channels, or whose branching
ratio into these channels is substantial. Here, the most obvious
example is provided by the decays
of light $K\mbox{-}$, $\eta$- and $\omega$-mesons into the three-pion final states, as well as by
the decays of heavier states, say,  $a_1(1260)\to\rho\pi\to 3\pi$ and
$a_1(1420)\to f_0(980)\pi\to 3\pi$. Further, the experimentally
observed decays of $X(3872)$ and $X(4260)$ mesons, which are candidates for
QCD exotica, 
are also largely dominated by three-particle final states. A well-known example
in the baryon sector is provided by the Roper resonance, which has a significant
branching ratio into the $\pi\pi N$ final state. A consistent calculation of the properties
of all above resonances thus implies an adequate treatment of the three-particle
sector on the lattice.

On the other hand, a proper handling of the three (and more) particle dynamics is a
necessary ingredient of the nuclear lattice theory beyond the calculation of spectra
and matrix elements of external currents for stable nuclei. For example, even the
reactions that involve the simplest nucleus -- the nucleon-deuteron scattering and the
breakup -- cannot be studied without setting a systematic theoretical framework for the
analysis of lattice data in the three-particle sector. In short, this framework should enable
one to ``translate'' the raw output from the calculations on the Euclidean lattices:
the two- and three-particle
energy levels, as well as the (real) matrix elements between different eigenstates of the
total Hamiltonian, into a bunch of the continuum observables: the binding energies,
amplitudes for the elastic scattering, rearrangement and breakup reactions, the mass and
width of various resonances as well as (generally complex) matrix elements of
external currents, etc. Given large number of continuum observables, it is not {\it a priori}
clear, how they can be extracted from the limited set of available lattice data.

A fundamental difference between lattice calculations and continuum studies consists in the
fact that lattice calculations are always done in a finite (Euclidean) volume, whereas
the study of the scattering processes in the continuum implies the definition of
asymptotic states, in which the incoming and outgoing particles travel over infinitely
large distances. Therefore, an infinite-volume limit in the scattering amplitudes, calculated
on a finite Euclidean lattice, cannot be straightforwardly performed -- this statement
can be related to the celebrated Maiani-Testa no-go theorem~\cite{Maiani:1990ca}.
L\"uscher has, however,
demonstrated that, in case of the two-body elastic scattering, it is possible to circumvent
this obstacle and extract the scattering phase directly from the finite-volume energy
spectrum in the two-particle sector~\cite{Luscher:1990ux}.
The infinite-volume limit can be then performed
numerically for the scattering phase, which contains only  exponentially suppressed
finite-volume corrections.

L\"uscher's finite-volume approach has become a standard procedure for the analysis
of lattice data in the two-body sector and has been successfully applied in a number
of recent studies (see, e.g.,~\cite{Briceno:2016mjc,Briceno:2017qmb,Moir:2016srx,Werner:2019hxc}). The approach has been further generalized for the
systems in the moving frames~\cite{Rummukainen,Gockeler:2012yj,Romero-Lopez:2018zyy},
and to the case of coupled two-body
channels (see, e.g., Refs.~\cite{Lage-KN,Lage-scalar,He,Sharpe,Briceno-multi,Liu,PengGuo-multi}). Moreover, as shown first by Lellouch and L\"uscher~\cite{Lellouch:2000pv}, very similar
methods can be used to treat two-body final-state interactions in decay processes.
Further developments of this idea can be found in Refs.~\cite{Kim:2005gf,Hansen:2012tf,Meyer:2011um,Bernard:2012bi,Agadjanov:2014kha,Briceno:2014uqa,Briceno:2015csa}. An alternative approach
to the study of
the scattering problems on the lattice is described in Ref.~\cite{Aoki:2013tba}.
A recent review on the present status of the scattering problem on the lattice
is given in Ref.~\cite{Briceno:2017max}.

It turns out, however, very difficult to directly generalize 
the ideas of L\"uscher's finite-volume method to the three-particle sector. The main
obstacle is related to the fact that, if more than two particles are present, they can form
compact
clusters, which are not interacting with each other and can be separated by large distances.
Putting differently, all interactions between particles do not necessarily vanish, if the size
of a system as a whole grows large. For this reason, it is not clear from the beginning,
whether the quantization condition (an equation that determines the energy levels of
a system in a finite three-dimensional box and can be obtained by using the boundary conditions on
the surface of the box) {\em does not} explicitly contain pair potentials that survive in the clusters. Were this not the case, the study
of the three-body spectrum would not provide any useful information about observables
of a system.

The aim of present contribution is to briefly review theoretical frameworks, which can be used
for the analysis of the finite-volume data in the three-particle sector. An emphasis will be
put on physical ideas, which lay the foundation of these approaches. It will be argued
that, apart from differences in technical details, all these approaches share the same basic
ideas and are equivalent to each other. I shall conclude by demonstrating few
examples of the data analysis, using the theoretical approaches described above.

\section{A brief history}
Historically, the first attempts to study three-particle system in a finite volume
has been undertaken within non-relativistic quantum
mechanics, using the Rayleigh-Schr\"odinger
perturbation theory for the ground-state energy level~\cite{Huang:1957im,Wu:1959zz}.
It turns out that the perturbation
expansion corresponds to the expansion of the energy shift in powers of $1/L$ (modulo
logarithms), where $L$ denotes the size of a box.
Recently, in
Refs.~\cite{Beane:2007qr,Tan:2007bg} and~\cite{Detmold:2008gh}, the
calculation of the ground-state energy shift was performed up to the order $L^{-6}$
and $L^{-7}$, respectively (in these calculations,
Refs.~\cite{Beane:2007qr,Detmold:2008gh} have used
non-relativistic EFT framework in dimensional regularization that allows one to achieve
the goal with a minimum effort). The result for the perturbative
shift of the energy level (due to the interactions between particles) up to order $L^{-5}$
is described in terms of a single observable quantity: the two-body scattering length $a$.
At order $L^{-6}$, two more parameters appear: the (observable)
two-body effective range $r$ and the genuine three-body coupling constant. The latter
can be also regarded as an observable since, in the perturbation theory, it can be related to
the (non-singular part of the) three-body scattering amplitude at threshold, which is
an observable. Thus, up to this order, the ground-state energy is determined only
by the infinite-volume observables in the three-particle sector. The above discussion
clearly shows the pattern of reasoning, which will be used in the following.

As already mentioned, the fundamental question,
which should be answered first and foremost, is, whether the on-shell data in the three-particle sector
(i.e., the $S$-matrix elements in the two- and three-particle sectors) are alone sufficient
for determining the finite-volume spectrum. The (rather complicated)
proof that this is indeed the case,
was first given in Ref.~\cite{Polejaeva:2012ut}.
The next big step forward was done in Refs.~\cite{Hansen:2014eka,Hansen:2015zga}, where the
quantization condition for three identical relativistic spinless particles has been derived
by summing up certain classes of Feynman diagrams in the three-particle Green function.
In Refs.~\cite{pang2} and~\cite{Mai:2017bge}, alternative derivations of the quantization
condition were presented, based on the non-relativistic effective field theory (NREFT) and
on the relativistic three-body unitarity, respectively. Despite very different starting points,
it is seen that all three approaches are equivalent modulo technical differences. Namely,
in all approaches, the finite-volume spectrum is characterized by the two-body input
(the on-shell two-body $K$-matrix) and the three-body input that carries different
names in different approaches
(the divergence-free threshold amplitude in Ref.~\cite{Hansen:2014eka,Hansen:2015zga},
the particle-dimer coupling constant in Ref.~\cite{pang2} and the
non-singular term which cannot be fixed by unitarity in Ref.~\cite{Mai:2017bge}), but
has the same physical content. The technical differences refer to the choice of the cutoff
functions, as well as the fact that the relativistic corrections are not (yet) included in the approach of Ref.~\cite{pang2} (this
project is underway). The equivalence of different approaches has been
recently discussed in Refs.~\cite{pang2,Jackura:2019bmu}.

Apart from the frameworks, which were listed above, one should also mention
Ref.~\cite{Briceno:2012rv}, where a
quantization condition was written down
by using the particle-dimer formalism very similarly to
Ref.~\cite{pang2}. In Refs.~\cite{Guo:2016fgl,Guo:2017ism,Guo:2018ibd},
alternative approach to the three-body problem in a finite volume,
based on the variational principle, has been suggested.
In Ref.~\cite{Aoki:2013kpa}, HAL QCD approach to the multi-particle systems has been
described. Finally, one has to specially mention earlier works~\cite{Kreuzer:2010ti,Kreuzer:2009jp,Kreuzer:2008bi,Kreuzer:2012sr}, where three-body equations were numerically
solved in a finite volume, and the dependence of the spectrum on various input parameters
has been thoroughly investigated. Three-nucleon systems were addressed within
Nuclear Lattice Effective Field Theory, see, e.g., Ref.~\cite{Lahde:2019npb}.

After the explicit form of the quantization condition in the three-particle sector
has been established, numerous papers have applied it to make predictions for 
the three-particle spectrum, as well as to perform analysis of the real
lattice data that are starting to appear. In addition, various technical issues were addressed and solved~\cite{Meissner:2014dea,Hansen:2015zta,Hansen:2016fzj,Hansen:2016ync,Briceno:2017tce,Sharpe:2017jej,pang1,Meng:2017jgx,Doring:2018xxx,Pang:2019dfe,Romero-Lopez:2018rcb,Mai:2019fba,Briceno:2018mlh,Briceno:2018aml,Mai:2018djl,Guo:2018xbv,Blanton:2019vdk,Romero-Lopez:2019qrt,Blanton:2019igq,Beane:2007es,Horz:2019rrn,Klos:2018sen,Konig:2017krd}. For the recent review, see Ref.~\cite{Hansen:2019nir} and references therein.

To summarize, after almost a decade of a concentrated effort by several groups, the
relation between the finite-volume three-particle spectrum on the lattice and the
parameters of the three-body $S$-matrix is now generally understood, and the formalism
has been already used to analyze the data. Three different frameworks are essentially
equivalent to each other, modulo technical issues. More complicated issues (e.g., an
analog of the Lellouch-L\"uscher formula for the three-particle decays) have not been
addressed yet.

\section{Essentials of the approach}

Below, I shall consider the derivation of the quantization condition by using the
non-re\-la\-ti\-vis\-tic effective Lagrangians which, in our opinion, provides
the most transparent setting for this purpose. One may also recall here a simple
and intuitive derivation of the two-body L\"uscher equation, which has been
performed in the same setting, see Ref.~\cite{Beane:2003yx}.
Further, as already mentioned, all derivations are essentially equivalent, albeit the use
of non-relativistic kinematics could be hard to justify when dealing with light particles
(especially, pions) on the lattice. This deficiency of the NREFT approach
has to be addressed at a later stage.

The final aim of lattice calculations is the extraction of the infinite-volume observables.
Since the calculations are always carried out in a finite volume, prior to the measurement,
one has to identify the quantities that have a smooth limit, when $L\to \infty$. The simplest
example of such quantity is the mass of a stable particle. The finite-volume corrections
to the masses are exponentially suppressed in $L$ and can thus be neglected, if $L$
is sufficiently large. Another example is given by the spacelike form factors of stable
particles. In general, each matrix element which, for a given kinematics, does not
develop an imaginary part in the infinite volume, receives exponentially suppressed
finite-volume corrections.

The situation changes, when one deals with the scattering problems. Consider,
for instance, the elastic scattering of two identical scalar particles. Two particles in the
intermediate state may go on shell in this kinematic region. In the infinite volume, this
leads to the non-vanishing imaginary part of the amplitude, which obeys elastic unitarity
below the first inelastic threshold. In a finite volume, on-shell particles can move far apart,
leading to the finite-volume effects that are no more exponentially suppressed (even,
are not suppressed at all). Hence, a question naturally arises, what are the quantities
that possess a smooth infinite-volume limit. The lattice calculations should aim at the
extraction of such quantities from data.

The problem has a simple and elegant solution within NREFT. The Lagrangian in this theory
(two-body sector) can be written down in the following form:
\eq\label{eq:L}
{\cal L}=\psi^\dagger\biggl(i\partial_0+\frac{\nabla^2}{2m}+\frac{\nabla^4}{8m^3}+\cdots\biggr)\psi-\frac{C_0}{2}\,\psi^\dagger\psi^\dagger\psi\psi+\frac{C_2}{4}\,(\psi^\dagger\stackrel{\leftrightarrow}{\nabla}^2\psi^\dagger\psi\psi+\mbox{h.c.})+\cdots\,.
\en
Here, $\psi(x)$ denotes the non-relativistic field with the mass $m$ and
$\stackrel{\leftrightarrow}{\nabla}=\frac{1}{2}\,
(\stackrel{\rightarrow}{\nabla}-\stackrel{\leftarrow}{\nabla})$. The propagator
of this field has the form
\eq
i\langle 0|T\psi(x)\psi^\dagger(y)|0\rangle=\int \frac{d^4p}{(2\pi)^4}\,
\frac{e^{-ip(x-y)}}{w({\bf p})-p_0-i\varepsilon}\,,\quad\quad
w({\bf p})=\frac{{\bf p}^2}{2m}\,.
\en
The relativistic corrections to the kinetic term ${\nabla^4}/{8m^3},\ldots$ are taken
into account perturbatively. In the following, in order not to overload the discussion with technicalities, I shall not consider these corrections.

The four-particle interactions in the Lagrangian are described by a tower of operators
with increasing mass dimension. Further, due to the structure of the non-relativistic
theory (a non-relativistic field has only one pole, corresponding to a particle, whereas the
antiparticle pole is absent), the structure of the perturbation series is remarkably simple.
Namely, only $s$-channel bubble diagrams contribute to the two-particle scattering amplitude. Restricting ourselves first to the lowest-order coupling $C_0$, the on-shell S-wave
scattering amplitude in the center-of-mass (CM) frame to
all orders in perturbation theory is given by:
\eq
T(E)=(-2C_0)+(-2C_0)\frac{1}{2}\,J(-2C_0)+(-2C_0)\frac{1}{2}\,J(-2C_0)\frac{1}{2}\,J(-2C_0)+\cdots\, ,
\en
where $E$ is the total energy in the CM frame $J$ corresponds to a single bubble:
\eq
J=\int \frac{d^D k}{(2\pi)^Di}\,\frac{1}
{(w({\bf k})-k_0-i\varepsilon)(w({\bf k})-E+k_0-i\varepsilon)}=\frac{imp}{4\pi}\, ,\quad\quad p=\sqrt{mE}\, .
\en
Dimensional regularization was used in the above expression to tame the ultraviolet
divergences.
Taking now into account all derivative terms and summing up the perturbative series,
one arrives at the following expression for $T(E)$:
\eq\label{eq:T}
T(E)=\frac{K(E)}{1-imp/(8\pi)\,K(E)}\, ,\quad\quad K(E)=(-2C_0)+(-2C_1)p^2+\cdots\,.
\en
Further, the elastic unitarity relates the quantity $K(E)$, which is the sum of all tree graphs,
to the elastic S-wave scattering phase:
\eq\label{eq:K}
K(E)=\frac{8\pi}{mp}\,\tan\delta(p)\,.
\en
Using the effective range expansion $p\cot\delta(p)=-1/a+\frac{1}{2}\,rp^2+O(p^4)$, it is a straightforward task to express the couplings $C_0,C_2,\ldots$ in terms of the physical observables $a,r,\ldots$.

An important remark is in order. Writing down the Lagrangian in Eq.~(\ref{eq:L}), I have consistently left out the so-called off-shell terms, i.e., the terms that can be eliminated by using equations of motion or field redefinitions. An example of such a term emerges first at order $\nabla^4$ and contains the operator
\eq
\hat O_{\mbox{\footnotesize off-shell}}=(\psi^\dagger\stackrel{\leftrightarrow}{\nabla}^4\psi^\dagger\psi\psi-
  \psi^\dagger \stackrel{\leftrightarrow}{\nabla}^2\psi^\dagger\psi\stackrel{\leftrightarrow}{\nabla}^2\psi)+\mbox{h.c.}\, .
  \en
  The tree-level contribution from this operator in the CM frame is proportional
  to $(p^2-q^2)^2$
  (here, $p$ and $q$ are the magnitudes of the relative 3-momenta in the incoming and
  outgoing states, respectively), and thus vanishes on the energy shell. Moreover,
  it can be seen that, inserting this vertex at any place in the bubble sum gives zero
  as well. In order to prove this, note that in this contribution $(p^2-q^2)^2$ will cancel
  one of the energy denominators, leading to the integral from a polynomial. As known, such integrals identically vanish in the dimensional regularization. And, of course, the final statement does not depend on the regularization used.

  Let us briefly summarize the properties of the NREFT approach, which will be important in the following. First of all, note that the term ``non-relativistic'' is used in two different
  contexts. The {\em kinematical} aspect was already mentioned above. It should be also
  mentioned that in the two-particle sector there exists a modified formulation of NREFT,
  in which the particles have relativistic dispersion law (i.e., the relativistic corrections in the
  one-particle propagators are summed up to all orders)~\cite{Colangelo:2006va,Gasser:2011ju}. Setting up a similar framework in case of three particles is in progress. From
  the {\em dynamical} standpoint, the term ``non-relativistic'' means the absence of
  explicit particle creation/annihilation processes in the theory -- all these processes are
  implicitly included into the effective couplings of the non-relativistic Lagrangian. Note
  that, in the relativistic theory, one should first single out the class of two-particle
  reducible diagrams, grouping all other diagrams into the kernels that have only exponentially vanishing dependence on $L$. This step is superfluous here since, owing to the particle number conservation in the NREFT, only two-particle reducible diagrams are left in the bubble sum, whereas irreducible kernels are parameterized with the low-energy polynomials, containing couplings from the effective Lagrangian. Putting differently, NREFT provides clean and efficient separation of the long-range (two-particle reducible diagrams) and short-range (the rest) effects, that makes the formalism very convenient for finite-volume calculations which are considered next.

  As first proposed in Ref.~\cite{Beane:2003yx}, the NREFT framework can be used
  to calculate the finite-volume spectrum of a system. The Feynman diagrammatic technique in a finite volume remains the same, except that all three-dimensional loop integrals
  over $d^3{\bf k}$ are replaced by the sums over discrete momenta ${\bf k}=2\pi{\bf n}/L\,,~{\bf n}=\mathbb{Z}^3$. The effective couplings, which describe short-range physics, are unaffected by the finite-volume effects (up to the exponentially suppressed terms).
  The finite-volume spectrum is given by the poles of the finite-volume analog of
  $T(E)$, given by Eq.~(\ref{eq:T}), where the infinite-volume expression $J=imp/(4\pi)$ is replaced by:
  \eq
  J_L=\int \frac{dk_0}{2\pi i}\frac{1}{L^3}\sum_{\bf k}
  \frac{1}{(w({\bf k})-k_0-i\varepsilon)(w({\bf k})-E+k_0-i\varepsilon)}
  =\frac{m}{2\pi^{3/2}L}\,Z_{00}(1;p_0^2)\, ,  
  \en
  where $p_0=pL/(2\pi)$ and $Z_{00}(1;p_0^2)$ stands for the so-called L\"uscher zeta-function. Note that dimensional regularization is implicit in the above expression -- otherwise,
  an infinite-volume term should be subtracted, which vanishes in dimensional regularization.

  The energy spectrum is then given by the equation:
  \eq\label{eq:luescher}
  p\cot\delta(p)=\frac{2}{\sqrt{\pi}L}\, Z_{00}(1;p_0^2)\, ,
  \en
  which is nothing but the L\"uscher equation (for simplicity, here
  I have not considered partial
  waves other than the S-wave, but these can be straightforwardly added to the Lagrangian and then appear in the equation).

  To summarize, in order to extract infinite-volume physics from lattice
  calculations, one has first parameterized ``good'' observables (i.e., the observables,
  in which the finite-volume effects are exponentially suppressed), in terms of the couplings
  of the NREFT Lagrangian, and then derived Eq.~(\ref{eq:luescher}), whose right-hand
  side is expressed trough $p_0^2$ that can be directly measured on the lattice. Moreover,
  the right-hand side has exponentially small corrections at large values of $L$ (because the left-hand side has this property).

  Two important remarks are in order. First, even if one started from the 
(non-relativistic)
 couplings, they are effectively summed up in the phase shift in the final result,
  according to Eq.~(\ref{eq:K}). This allows one not to worry about the convergence of the
  effective-range expansion and deal directly with the observable $\delta(p)$
  at a measured value of $p$. Such a luxury, however, is in general
  no more affordable in
  complicated systems (e.g., the ones, containing three particles).
  For example, an attempt of the resummation might lead to the cutoff-dependent
  unconventional
  amplitudes, like the ones introduced in Refs.~\cite{Hansen:2014eka,Hansen:2015zga}. The separation of the short- and long-range physics with the short-range physics encoded in the couplings, which lies in the basis of the NREFT approach is, however, universal. For this reason, I do not try to do any resummation in the three-particle sector
  and identify the low-energy
  couplings as {\em the observables} that should be extracted from lattice data.

  The second remark concerns the role of the off-shell terms in a finite volume. As seen
  above, such terms do not contribute to the on-shell amplitude in the infinite volume, because the insertion of the off-shell vertices into the diagrams leads to the no-scale integrals that vanish in dimensional regularization. Exactly the same mechanism works in a finite volume, and the no-scale integrals lead to the vanishing sums also here. Hence, the quantization condition (the L\"uscher equation) does not depend on the off-shell couplings, meaning that the finite-volume spectrum is determined solely by the on-shell input in the infinite volume.

  After discussing the two-body case in detail, I wish to turn to the three-particle case, which
  is our ultimate goal. It is seen that, within NREFT, the derivation follows the same pattern.
  The non-relativistic Lagrangian is amended by tower of terms that describe 
  six-particle interactions:
  \eq
  {\cal L}_3=-\frac{D_0}{6}\, \psi^\dagger\psi^\dagger\psi^\dagger\psi\psi\psi
  -\frac{D_2}{12}\,(\psi^\dagger\psi^\dagger \nabla^2\psi^\dagger\psi\psi\psi
  +\mbox{h.c.})+\cdots\,.
  \en
Here, I assume that the three particles are in the rest frame, so there
is no need to introduce the Galilei-invariant derivative $\stackrel{\leftrightarrow}{\nabla}$.
  As in the two-particle case, the off-shell terms are left out, for the same reason. Inclusion
  of such terms give rise to the no-scale integrands which (after integration/summation) vanish both in the infinite and
  in a finite volume. The details of the proof are given in Ref.~\cite{pang2}, but the
  statement seems almost obvious. Indeed, one knows from the beginning that, in the infinite volume, the off-shell terms do not contribute to the $S$-matrix. This may happen,
  if and only if all insertions of the off-shell vertices vanish, i.e., lead to the no-scale integrals. Then, the corresponding insertions
  in a finite volume should vanish as well, since the structure of the diagrams is the same.
  Thus, a complicated proof, given first in Ref.~\cite{Polejaeva:2012ut}, becomes almost trivial in the NREFT approach!

  At the next step, it is convenient to introduce a fictitious dimer field in the theory,
  replacing two-particle interactions by the propagation of a dressed dimer field
  (for details, see, e.g., Ref.~\cite{Bedaque:1998kg}). This is a purely mathematical
  trick that is equivalent to the introduction of a new variable in the path integral, defining the generating functional of the theory. However, this trick greatly simplifies the bookkeeping in the three-particle case, and I shall use it in the following. I would like to stress
  from the beginning that using particle-dimer picture does not mean assuming the existence of a stable dimer (or a narrow resonance) with given quantum numbers, so the proposed
  approach is completely general.

  The Lagrangian in the particle-dimer picture can be rewritten in the following manner:
  \eq
  {\cal L}=\psi^\dagger\biggl(i\partial_0+\frac{\nabla^2}{2m}\biggr)\psi
  +\sigma T^\dagger T+\frac{1}{2}\,f_0(T^\dagger\psi\psi+\mbox{h.c.})
  +h_0T^\dagger T\psi^\dagger\psi+\cdots\,.
  \en
  Here, $T$ stands for the dimer field, $\sigma=\pm 1$ is chosen to match the sign of $C_0$, and the dots stand for the higher-order terms with derivatives. Note also that here
  I have displayed the S-wave dimer only, albeit the formalism allows for the introduction of a dimer with arbitrary spin (see Ref.~\cite{pang2} for more detail). Further, in order to be equivalent to the original theory, the coupling $f_0,h_0$ constants should obey
  the relations:
  \eq
  f_0^2=2C_0\sigma\, ,\quad\quad
  3f_0^2h_0=-2D_0\, .
  \en
  These relations can be easily proved by using the equations of motion for the field $T$ in
  the Lagrangian.

  Further, the dressed propagator of a dimer field
  in the CM frame corresponds to the sum over
  all bubble diagrams in the two-particle sector:
  \eq
  D(E)  =\frac{Z_d}{p\cot\delta(p)-ip}\, ,\quad\quad p=\sqrt{mE}\, ,
  \en
  where the constant $Z_d$ does not depend on $E$. The expression $p\cot\delta$ is obtained
  by summing up all interaction vertices with arbitrary number of derivatives,
in the two-body sector. At lowest order,
  when only the term with $f_0$ is present, $p\cot\delta(p)=-1/a$.

\begin{figure}[t]
    \begin{center}
      \includegraphics*[width=15.cm]{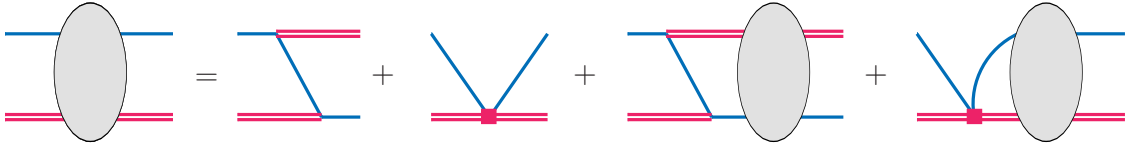}
      \caption{Bethe-Salpeter equation for the particle-dimer scattering amplitude. Double line and the shaded box stand for the dimer propagator and the particle-dimer coupling $h_0$, respectively.}
      \label{fig:BS}
    \end{center}
    \end{figure}

  The particle-dimer scattering amplitude obeys the Bethe-Salpeter equation:
  \eq
  {\cal M}({\bf p},{\bf q};E)=Z({\bf p},{\bf q};E)
  +8\pi\int^\Lambda\frac{d^3{\bf k}}{(2\pi)^3}\,
  Z({\bf p},{\bf k};E)\tau({\bf k};E){\cal M}({\bf k},{\bf q};E)\,,
  \en
  where
  \eq
  \tau({\bf k};E)=\frac{1}{k^*\cot\delta(k^*)-ik^*}\, ,\quad\quad
  -ik^*=\sqrt{\frac{3}{4}\,{\bf k}^2-mE}\, ,
  \en
  and
  \eq
  Z({\bf p},{\bf q};E)=\frac{1}{-mE+{\bf p}^2+{\bf q}^2+{\bf p}{\bf q}}
  +\frac{h_0}{mf_0^2}+\cdots\, .
  \en
  The ellipses above stand for the terms with more derivatives and $\Lambda$ denotes
  the ultraviolet cutoff. The couplings $h_0,\ldots$ (but not $f_0,\ldots$ ) depend on $\Lambda$, so that
  the amplitude ${\cal M}$ is $\Lambda$-independent.
  Schematically, the Bethe-Salpeter equation is shown in Fig.~\ref{fig:BS}. Note also
  that the three-particle scattering amplitude can be expressed through the particle-dimer
  scattering amplitude by merely attaching particle-dimer vertices to the external dimer
  lines.

    Acting now in the same way as in the two-particle case, one can straightforwardly
    rewrite the equation
    in a finite volume. The integral over $d^3{\bf k}$ gets discretized, and $ip^*$ gets
    replaced by the L\"uscher zeta function. The resulting equation takes the form:
\eq\label{eq:ML}
{\cal M}_L({\bf p},{\bf q};E)=Z({\bf p},{\bf q};E)
  +\frac{8\pi}{L^3}\,\sum_{\bf k}^\Lambda
  Z({\bf p},{\bf k};E)\tau_L({\bf k};E){\cal M}_L({\bf k},{\bf q};E)\,,
  \en
  where
  \eq\label{eq:tauL}
  \tau_L^{-1}({\bf k};E)=k^*\cot\delta(k^*)-\frac{4\pi}{L^3}\sum_{\bf l}
  \frac{1}{{\bf k}^2+{\bf l}^2+{\bf k}{\bf l}-mE}\,.
  \en
  Note that in Eq.~(\ref{eq:tauL}), unlike Eq.~(\ref{eq:ML}), dimensional regularization
  is used to tame the ultraviolet divergence.

  The finite-volume spectrum is determined by the poles of the three-particle scattering amplitude, which can be expressed through the particle-dimer scattering amplitude in the same way as in the infinite volume. Further, Eq.~(\ref{eq:ML}) represents a system of linear
  equations in the basis of three-momenta. The amplitude develops poles, when the determinant of the system vanishes. 
It can be shown~\cite{Doring:2018xxx} that the three-particle quantization condition follows:
  \eq
  \mbox{det}\biggl(\delta_{{\bf p}{\bf q}} \tau_L^{-1}({\bf p};E)
  -\frac{8\pi}{L^3}\, Z({\bf p},{\bf q};E)\biggr)=0\, .
  \en
  The values of $E$, which are the solutions of this equation, determine the spectrum
  of three identical particles in a finite volume.

  This brings us to the strategy for the extraction of the physical observables on the lattice.
  The lattice calculations provide us with discrete values $E=E_n$ in the three-particle sector. In addition, measurements in the two-particle sector should be performed. On the other
  hand, the quantization condition parameterizes the lattice spectrum in terms of ``good''
  observables $p\cot\delta(p)$ (which contains the whole bunch of the two-body couplings $C_0,C_2,\ldots$ summed up) and the particle-dimer couplings $h_0,\cdots$ (these correspond
  to the genuine three-body force). Then, the workflow in the data analysis
  might look as follows:

  \begin{itemize}
  \item First, one determines the two-body scattering phase from the measurements
    in the two-body sector alone. In the analysis of the three-body data, the scattering phase
    is fixed.
  
  \item
    Next, determine the three-body couplings $h_0,\ldots$ from the fit to the data in the
    three-particle sector.

  \item
    The couplings $h_0,\ldots$ are ``good'' observables in the sense that they contain only
    exponentially suppressed finite-volume effects. They are also observables in the sense
    that they can be unambiguously related to the infinite-volume $S$-matrix elements
    (i.e., all off-shell operators, which would be redundant, were already eliminated from the theory). In order to extract the observables one is interested in (the amplitudes for
    various reactions in the three-particle sector), one should solve the Bethe-Salpeter
    equation in the infinite volume, using the values of the couplings $h_0,\ldots$
    that were previously determined on the lattice.

  \end{itemize}

  The procedure, which was described above, solves the problem of extraction of the three-particle observables from the lattice data in principle. The procedure is very similar
  to the one used in the two-body sector, with one difference. In the two-body sector,
  one has an algebraic relation between the scattering phase and the effective couplings
  that can be established to all orders. In the three-particle sector, this property is lost,
  and the calculations should be done at a given order in the low-energy expansion.
The couplings and the amplitudes are now related through an integral equation.

From the above derivation it is seen that using the framework of NREFT
allows one to achieve the goal with a surprising ease and elegance. 

\section{Analysis of the lattice data}

The quantization condition, which was derived above (as well as alternative formulations,
see Refs.~\cite{Hansen:2014eka,Hansen:2015zga} and~\cite{Mai:2017bge}), can be
further generalized to include coupling of two- and three-body
channels~\cite{Briceno:2017tce} as well as subthreshold resonances and bound states in the two-particle sub-systems~\cite{Briceno:2018aml,Romero-Lopez:2019qrt}. Higher partial waves can be included from the beginning~\cite{Hansen:2014eka,pang2}. Using the
symmetries of the cubic lattice, the quantization condition can be projected onto various
irreducible representations of the cubic group~\cite{Doring:2018xxx,Blanton:2019igq}. The quantization condition can be solved by using a simple input, and the qualitative behavior of the energy levels in the three-particle system can be studied, where one observes an interesting phenomenon of repulsion of the three-particle and particle-dimer levels that leads to the avoided level crossing. The signatures of the three-body resonances can also be studied~\cite{pang2,Doring:2018xxx,Klos:2018sen,Briceno:2018mlh,Blanton:2019igq,Romero-Lopez:2019qrt}. Further, the finite-volume shift of the three-body shallow bound states can be derived~\cite{Meissner:2014dea,Hansen:2016ync,pang1}, etc.

Below, I would like to briefly dwell on the use of the formalism to analyze the data on the
shift of the three-particle and particle-dimer energy levels. The quantization condition
admits a systematic perturbative expansion in powers of $1/L$ (modulo logarithms) in the
vicinity of free finite-volume energies. The result for the ground state of the two- and
three-particle energy levels reads~\cite{Hansen:2015zta,Pang:2019dfe}:
\eq\label{eq:E23}
E_2&=&2m+\frac{4\pi a}{mL^3}\,\biggl(1+c_1\biggl(\frac{a}{\pi L}\biggr)
+c_2\biggl(\frac{a}{\pi L}\biggr)^2
+c_3\biggl(\frac{a}{\pi L}\biggr)^3
+\frac{2\pi ra^2}{L^3}-\frac{\pi a}{m^2L^3}\biggr)+O(L^{-7})\, ,
\nonumber\\[2mm]
E_3&=&3m+\frac{12\pi a}{mL^3}\,\biggl(1+d_1\biggl(\frac{a}{\pi L}\biggr)
+d_2\biggl(\frac{a}{\pi L}\biggr)^2
+\frac{3\pi a}{m^2L^3}+\frac{6\pi ra^2}{L^3}
+d_3\biggl(\frac{a}{\pi L}\biggr)^3\ln\frac{mL}{2\pi}\biggr)
\nonumber\\[2mm]
&-&\frac{D}{48m^2L^6}+O(L^{-7})\,,
\en
where $c_i,d_i$ are known coefficients and $D$ contains the three-body force
(for instance, the coupling constant $h_0$, in the NREFT language, or the divergence-free
threshold amplitude, in the language of Refs.~\cite{Hansen:2014eka,Hansen:2015zga}).
The expressions for
the energy shift of the lowest excited states in various irreducible representations
of the cubic group, as well as for the leading-order shift of the particle-dimer bound state
are also available~\cite{Pang:2019dfe}. The results are in agreement with those from
Ref.~\cite{Beane:2007qr}. Note also that the terms proportional to $1/m^2$ are
relativistic corrections and are absent in the non-relativistic framework.
Note also that the similar formulae, derived with the use of Rayleigh-Schr\"odinger perturbation theory, exist for the $n$-particle states as 
well~\cite{Huang:1957im,Wu:1959zz,Beane:2007qr,Tan:2007bg,Detmold:2008gh}.

As seen from the above expressions, the two-body scattering length $a$ and the
effective range $r$ can be obtained from the fit of the $L$-dependence of the
two-body spectrum alone. What can be extracted in addition from the three-body spectrum
up-to-and-including $O(L^{-6})$,
is only the three-body coupling $h_0$ which is related to $D$.
All other observables can be then predicted
in principle, using the determined value of $h_0$ in the infinite-volume equations
(of course, one could alternatively perform the global fit of the
parameters $a,r,D$ to the $E_2,E_3$). In our opinion, at present, this is the most efficient
way the handle the data from the three-particle sector.

\begin{figure}[t]
 \begin{center}
    \includegraphics*[width=5.6cm]{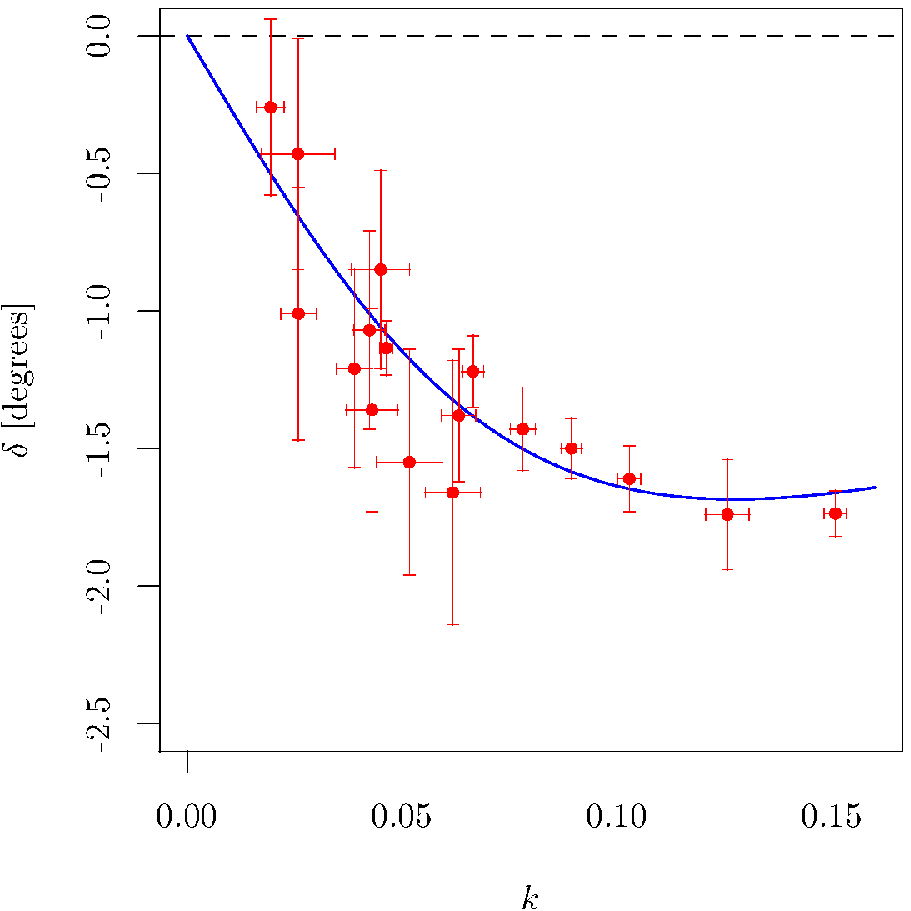}\hspace*{2.cm}
     \includegraphics*[width=5.6cm]{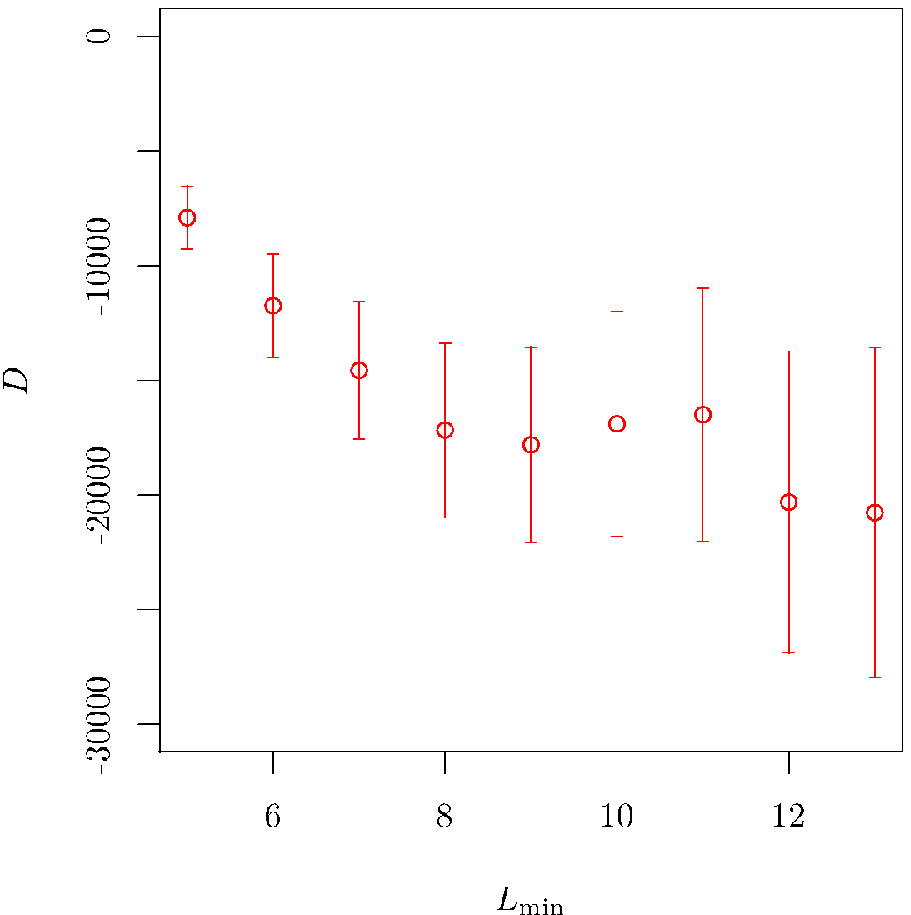}
   \end{center}
\vspace*{-.4cm}
   
   \caption{The two-body S-wave phase shift (left panel) and the three-body coupling $D$
     for different fit ranges from $L_{min}\leq L\leq L_{max}$ (right panel).  $L_{max}=24$ is fixed
     and  $L_{min}$ is varied.}\label{fig:Romero}
   \end{figure}

In the last years, several attempts have been undertaken to extract the infinite-volume
information on the lattice.  
In Ref.~\cite{Beane:2007es}, the ground-state energy levels in the systems with
up to five charged pions were fitted with the use of the perturbative formula for different
pion masses. This has resulted in the three-body coupling, significantly different from zero
for smaller pion masses. The same data has been fitted in Ref.~\cite{Mai:2018djl}
by using the framework of Ref.~\cite{Mai:2017bge}, and
the counterpart of the three-body force has been determined. 
In Ref.~\cite{Romero-Lopez:2018rcb}, an exploratory study of the extraction of the
three-body force in the $\varphi^4$ theory has been performed. The lattice
calculations were performed at many different values of $L$ that enables one to
study the dependence of the extracted parameters on the choice of the fit range.
The results of this investigation are shown in Fig.~\ref{fig:Romero}, where it is seen
that the parameter $D$ is non-vanishing at $4\sigma$ for the optimum fit range with
$L_{min}=9$. Finally, in Ref.~\cite{Horz:2019rrn}, the two- and three-particle
ground and excited levels energy levels for charged pions have been calculated and analyzed,
resulting in the non-zero value of the three-body threshold amplitude. The two-body scattering length and the effective radius were determined as well. The data from  Ref.~\cite{Horz:2019rrn} have been recently analyzed in Refs.~\cite{Mai:2019fba,Blanton:2019vdk}.

   It should be pointed out that the main obstacle in the analysis of the three-particle data
   consists in the fact that the three-body force comes first at N$^3$LO in the $1/L$
   expansion. This suggests to perform the fit at not so large values of $L$. However, in
   this case, the exponentially suppressed effects can become non-negligible, affecting
   the systematic error in the extraction. This issue was briefly addressed in
   Ref.~\cite{Romero-Lopez:2018rcb}. It was namely demonstrated that, at lowest order
   in perturbation theory, using the volume-dependent mass $m(L)$ instead of the
   infinite-volume mass $m$ in Eq.~(\ref{eq:E23}) accounts for the leading exponential
   effect. Since the $\varphi^4$ theory for a given set of input parameters seems to be
   perturbative to a good approximation (see Fig.~\ref{fig:Romero}, left panel), this
   trick might help to substantially reduce the systematic error in the analysis. The same
   statement will be true in the meson ChPT.
   It remains to be seen, whether the result persists in higher orders in perturbation theory.

   \section{Summary and outlook}

   Below, I briefly summarize the recent progress in the study of the three-particle
   system on the lattice and suggest the issues that could be addressed in the
   near future.

   \begin{itemize}

   \item
     Owing to the consolidated effort of several research groups, the relation between the
     three-particle energy spectrum and three-particle observables in the infinite volume is
     now firmly established that provides a solid framework for the analysis of data
     in the three-particle sector. Adjusting the framework for the analysis of concrete
     systems (say, to the case of the Roper resonance) will still require some effort, regarding the technical details. 

   \item
     At present, the use of the perturbative expansion for the ground and excited state energy levels has proven to be the optimum strategy for the extraction of the three-body force
     from lattice data. The errors, however, are still substantial, owing largely
     to the fact that the three-body
     force emerges first at order $L^{-6}$. Carrying out simulations at smaller values
     of $L$ in order to enhance its relative weight, one should make sure that the systematic uncertainties, caused by the exponentially suppressed polarization effects, are under control. A global fit, using the  ground-state and excited states, particle-dimer levels, as well as input from the multiparticle energy levels may be of great use to reduce the uncertainty.

   \item
     The study of the three-particle decays (a derivation of an analog to the Lellouch-L\"uscher formula in the three-particle case) remains an immediate objective.

     \end{itemize}

     {\em Acknowledgments:} The author
     acknowledges the support from the DFG
(CRC 110 ``Symmetries and the Emergence of Structure in
QCD''), as well as from
Volkswagenstiftung under contract no. 93562.
The work was in part supported by Shota Rustaveli National Science Foundation
(SRNSF), grant no. DI-2016-26.

\end{document}